\documentclass[conference,10pt]{IEEEtran}
\usepackage{epsfig,epsf,rotating,setspace,latexsym,amsmath,amssymb,amsfonts,bm,theorem,subfigure,epstopdf,cite,authblk,bbm}
\usepackage{algorithm}
\usepackage[noend]{algpseudocode}
\usepackage{color}
\usepackage{mathtools}
\usepackage{soul}

\newcommand{\secref}{Section~\ref}
\newcommand{\lemmaref}{Lemma~\ref}
\newcommand{\figref}{Fig.~\ref}

\algrenewcommand\algorithmicforall{\textbf{foreach}}
\algrenewcommand\algorithmicindent{.8em}

\newtheorem{theorem}{Theorem}
\newtheorem{lemma}{Lemma}

\newtheorem{remark}{Remark}

\newenvironment{Proof}[1]{\medskip\par\noindent{\bf Proof:\,}\,#1}{{\mbox{\,$\blacksquare$}\par}}

\IEEEoverridecommandlockouts
\allowdisplaybreaks

\begin{document}

\title{Age of Gossip on a Grid}
 
\author{Arunabh Srivastava \qquad Sennur Ulukus\\
        \normalsize Department of Electrical and Computer Engineering\\
        \normalsize University of Maryland, College Park, MD 20742\\
        \normalsize  \emph{arunabh@umd.edu} \qquad \emph{ulukus@umd.edu}}

\maketitle

\begin{abstract}
    We consider a gossip network consisting of a source generating updates and $n$ nodes connected in a two-dimensional square grid. The source keeps updates of a process, that might be generated or observed, and shares them with the grid network. The nodes in the grid network communicate with their neighbors and disseminate these version updates using a push-style gossip strategy. We use the version age metric to quantify the timeliness of information at the nodes. We find an upper bound for the average version age for a set of nodes in a general network. Using this, we show that the average version age at a node scales as $O(n^{\frac{1}{3}})$ in a grid network. Prior to our work, it has been known that when $n$ nodes are connected on a ring the version age scales as $O(n^{\frac{1}{2}})$, and when they are connected on a fully-connected graph the version age scales as $O(\log n)$. Ours is the first work to show an age scaling result for a connectivity structure other than the ring and fully-connected networks that represent two extremes of network connectivity. Our work shows that higher connectivity on a grid compared to a ring lowers the age experience of each node from $O(n^{\frac{1}{2}})$ to $O(n^{\frac{1}{3}})$.
\end{abstract}

\section{Introduction}
With the roll-out of 5G communication, and a focus on decentralized wireless communication systems in 6G, including seamless machine-machine and human-machine interactions for time-critical tasks, timeliness has become an important metric to consider. This is of importance in many upcoming applications, such as drone swarms, networks of self-driving cars, and remote medical procedures. It is now known that latency alone is not sufficient to characterize timeliness \cite{popovski2022perspective}, and new metrics are needed. One such metric is age of information \cite{kaul2012real,sun2019age, yatesJSACsurvey}. Several related freshness metrics are age of incorrect information \cite{maatouk20AOII}, age of synchronization \cite{zhong18AoSync}, binary freshness metric \cite{cho3BinaryFreshness}, and version age of information \cite{yates21gossip, Abolhassani21version, melih2020infocom}.

In this paper, we consider the version age of information metric. In simple words, the version age of information of a node in a network is the number of versions behind the node is with respect to the source node which has the latest version of the update. Reference \cite{yates21gossip} characterizes the version age of information for general networks with Poisson update times and describes a set of recursive equations that can be used to calculate the average version age of connected subsets of a network. Reference \cite{yates21gossip} also shows that the average version age of a single node in a fully-connected network scales as $O(\log{n})$ with the network size $n$. Reference \cite{yates21gossip} experimentally observes, and reference \cite{buyukates21CommunityStructure} mathematically shows, that for a ring network, the version age scales as $O(\sqrt{n})$. Reference \cite{kaswan22jamming} shows that the average version age in a line network scales the same as the version age in the corresponding ring network, as the line and ring networks differ only at a missing link connecting both ends of the line. 

\begin{figure}[t]
    \centering
    \includegraphics[width = \linewidth]{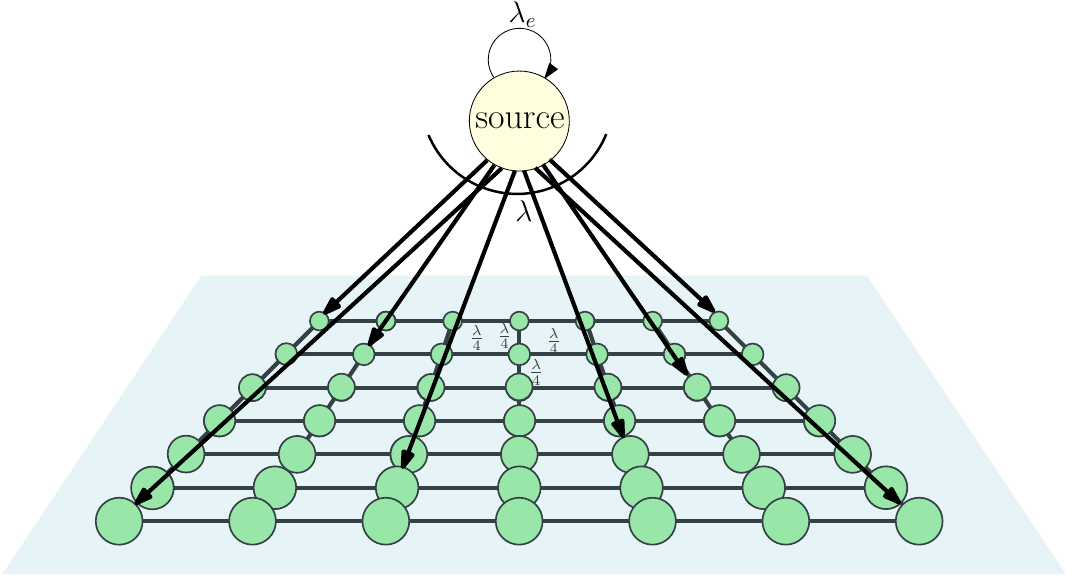}
    \caption{A gossip network where the node in light yellow is the source generating updates and sending them to a network of nodes connected in a grid. There are no boundaries in the grid, and all nodes have four neighbors.}
    \label{fig1}
\end{figure}

Our work here extends these results by considering a gossip network connected in the form of a two-dimensional square grid; see \figref{fig1}. This network has a higher connectivity than the ring (also the line) network as each node in a grid is connected to its four neighbors whereas each node on a ring is connected to its two neighbors, and has a lower connectivity than the fully-connected network as each node in a fully-connected network is connected to all of the remaining $n-1$ nodes. In this paper, we show that the version age of a node in a two-dimensional square grid scales as $O(n^{\frac{1}{3}})$. This result supports that increased connectivity decreases version age.

\section{System Model and the Version Age Metric}

We consider a wireless system which has one source node generating or observing updates of a particular process. This source updates as a rate $\lambda_e$ Poisson process independent of all other processes in the network. The system also has a gossip network, denoted by $\mathcal{N}$, that consists of $n$ nodes. The source node shares the updates with the gossip network as a Poisson process with a total rate $\lambda$, giving each node in the network an equal chance of being updated. This can be understood alternatively as, the source sends updates to each node as a Poisson process of rate $\frac{\lambda}{n}$ independent of all other processes.

\begin{figure}[t]
    \centering
    \includegraphics[width = 0.8\linewidth]{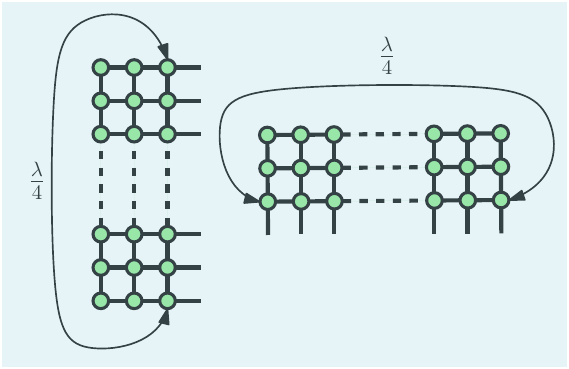}
    \caption{All nodes have four connections. Nodes that are seemingly at the boundary in \figref{fig1} are connected in a wrap-around fashion to the nodes on the opposite side of the row or column.}
    \label{fig_cross_connections}
\end{figure}

\begin{figure}[t]
    \centering
    \includegraphics[width = 0.55\linewidth]{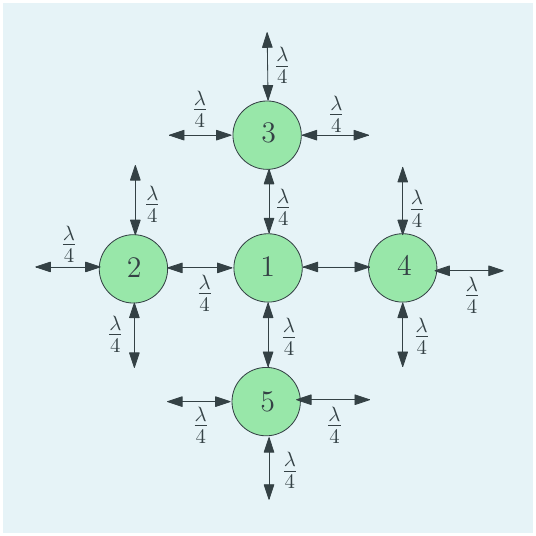}
    \caption{Each node updates its four neighbors with an update ate of $\frac{\lambda}{4}$. Total update rate per node is $\lambda$.}
    \label{fig2}
\end{figure}

The network of gossiping nodes is arranged in a two-dimensional square grid structure. Each node in the grid has four neighbors, in the sense that there are no boundaries in the network, i.e., a  node that is seemingly at the boundary in \figref{fig1} is connected in a wrap-around fashion to the nodes on the opposite side of the row or column as shown in \figref{fig_cross_connections}. Each node has a total gossiping rate of $\lambda$ which it divides into four equal parts to gossip with its neighbors. Hence, each node in the grid gossips with its four neighbors in a push-style gossiping protocol as a rate $\frac{\lambda}{4}$ Poisson process, as shown in \figref{fig2}, independent of all other processes in the network.

To quantify the freshness of update versions at the nodes, we use the version age metric. We start by defining the counting process associated with the updates at the source as $N_s(t)$. For node $i$ in the gossip network, we define the counting process associated with the updates as $N_i(t)$. Next, we define the version age of node $i$ as $X_i(t) = N_s(t) - N_i(t)$. Hence, the version age of any node in the grid network is the number of versions the particular node is behind with respect to the source node. Similarly, we define the version age of a connected subset $S$ of the grid as $X_S(t) = \min_{j \in S} X_j(t)$. Note that $X_S(t)$ is the smallest version age in the subset of nodes in $S$. Finally, we define the limiting average version age of set $S$ as $v_S = \lim_{t \rightarrow \infty} \mathbb{E}[X_S(t)]$. 

The version age of a node evolves as follows: If the source node updates itself, then the version age of every node in the gossip network increases by $1$. If the source node sends an update to a node, then the node's version age reduces to $0$. If a node $i$ sends its version of the update to node $j$, then $j$ updates itself if it has an older version than the one sent by $i$. Otherwise, it keeps its own version and rejects $i$'s version. 

We define the rate of information flow from node $i$ to node $j$ as $\lambda_{ij}$ which is the rate of the Poisson update process from $i$ to $j$. We say that node $i$ is a neighboring node of set $S$ if $\lambda_{ij} > 0$ for some $j \in S$. We define the total rate of information flow into the set $S$ from neighboring node $i$ as $\lambda_i(S)$, 
\begin{align}
    \lambda_i(S) = \begin{cases}
        \sum_{j \in S} \lambda_{ij}, & i \notin S\\
        0, & \text{otherwise}
    \end{cases}
\end{align}
Similarly, $\lambda_0(S)$ is the total rate of information flow from the source node into the set $S$. Finally, we define $N(S)$ to be the set of nodes which are neighboring nodes to the set $S$. With these definitions, we are now equipped to use the recursive equations for general networks proposed in \cite{yates21gossip}.

\section{Version Age in a Grid}\label{sec3}

In \cite{yates21gossip}, Yates developed a recursive formula to find the version age of a node in a general network as,
\begin{align}
    v_S = \frac{\lambda_e + \sum_{i \in N(S)}\lambda_i(S)v_{S \cup \{i\}}}{\lambda_0(S) + \sum_{i \in N(S)}\lambda_i(S)} \label{yates-recursion}
\end{align}
This recursive formula expresses the version age of a subset $S$ as a linear combination of the version ages of subsets that are obtained by adding one more neighboring node to this subset, i.e., $S \cup \{i\}$. For a general network, the number of such equations obtained is exponential in the number of nodes, and is difficult to solve in closed-form.

We want to find the version age of a typical node $j$, $v_{\{j\}}$ in a grid network. From the symmetry of the wrapped-around grid network, all nodes will have the same version age. Towards that end, we need to start with a single node $j$, i.e., $S=\{j\}$ in \eqref{yates-recursion}, and keep adding neighboring nodes $i$ until the entire network is exhausted. We know that the version age of the entire network $v_{\mathcal{N}}$ is  $v_{\mathcal{N}}=\frac{\lambda_e}{\lambda}$. In principle, working backwards, we can find the version age of node $j$, $v_{\{j\}}$, by solving all of the linear equations obtained through the recursion in \eqref{yates-recursion}. To achieve that, we need to identify and exploit simplifications/symmetries in these equations.

References \cite{yates21gossip} and \cite{buyukates21CommunityStructure} were able to find the version ages of fully-connected and ring networks, respectively, by developing relatively simple bounds, as in those cases, the age of a subset depends only on the size of the subset due to the symmetry of the networks. However, in the case of a grid network, subsets of the same size may have widely varying ages. Consider for instance a subset of $9$ nodes: those $9$ nodes could be on a line, versus on a $3\times 3$ square, versus on an $L$ shape with $6$ nodes on the longer arm of $L$ and $4$ nodes on the shorter arm. All these subsets of $9$ nodes, even in a symmetric wrapped-around grid network, will have different version ages.

To tackle this problem, in \secref{sec3_1}, we find a general upper bound (Lemma~\ref{lemma1}) and a general lower bound (Lemma~\ref{lemma2}) for $v_S$ in a general network. In \secref{sec3_2}, we find an upper bound for $v_S$ for the specific grid network (Lemma~\ref{grid_recursion}). Then, we prove \lemmaref{floor_inequality}, which is a mathematical inequality, to be used for proving our main result. Finally, we prove the main result of our paper, Theorem~\ref{thm}, which states that the version age of a single user scales as $O(n^{\frac{1}{3}})$ in a grid network.

\subsection{Upper and Lower Bounds for $v_S$ in a  General Network} \label{sec3_1}

The recursion in \eqref{yates-recursion} provides a method to find the version age of any connected subgraph of a gossip network. Even for the simplest graphs, such as rings and and fully-connected networks, the application of this recursion is tedious. For more complex graphs, such as the two-dimensional grid, use of symmetry is restricted and direct application of the recursion is challenging since the number of sets grows quickly. In order to simplify the application of the recursion, the following lemma modifies the recursion to find bounds for sets only based on the one-expanded set that has the highest version age as opposed to all one-expanded sets. This enables us to use the geometry of the network and restrict the number of sets we need to handle to find a bound for the set in question.

\begin{lemma} \label{lemma1}
    In any general gossip network, the following upper bound holds for a subset of nodes $S$,
    \begin{align}
        v_S \leq \frac{\lambda_e + |N(S)| \min_{i\in N(S)}{\lambda_i(S)} \max_{i\in N(S)}{v_{S \cup \{i\}}}}{\lambda_0(S) + |N(S)|\min_{i\in N(S)}\lambda_i(S)}  \label{eq_ub}
    \end{align}
\end{lemma}

\begin{Proof}
    We start by rearranging \eqref{yates-recursion} as follows,
\begin{align}
     \lambda_e = \lambda_0(S)v_S + \sum_{i \in N(S)}\lambda_i(S)(v_S - v_{S \cup \{i\}})  \label{rearranged_recursion}
\end{align}
Now, we lower bound the sum on the right hand side as,
\begin{align}
    \!\!\!\lambda_e &\geq \lambda_0(S)v_S + |N(S)|\min_{i \in N(S)}\lambda_i(S)(v_S \!-\! v_{S \cup \{i\}})\\
    &\geq \lambda_0(S)v_S + |N(S)|\min_{i\in N(S)}\lambda_i(S) \min_{i \in N(S)}(v_S \!-\! v_{S \cup \{i\}}) \! \\
    &= \lambda_0(S)v_S + |N(S)|\min_{i \in N(S)}\lambda_i(S)(v_S \!-\! \max_{i \in N(S)} v_{S \cup \{i\}}) \! \label{eq7}
\end{align}
Rearranging \eqref{eq7} gives the desired result.
\end{Proof}

In a similar way, we can also prove the lower bound in the following lemma.

\begin{lemma} \label{lemma2}
    In any general gossip network, the following lower bound holds for a subset of nodes $S$,
    \begin{align}
        v_S \geq \frac{\lambda_e + |N(S)| \max_{i\in N(S)}{\lambda_i(S)} \min_{i\in N(S)}{v_{S \cup \{i\}}}}{\lambda_0(S) + |N(S)|\max_{i\in N(S)}\lambda_i(S)}\label{eq_lb}
    \end{align}
\end{lemma}

\subsection{An Upper Bound for $v_S$ in a Grid}\label{sec3_2}
We now modify the recursion in \eqref{yates-recursion} of \cite{yates21gossip} in a different way, suited to the two-dimensional grid network. We will follow the technique that we used in \secref{sec3_1}. 

\begin{lemma}\label{grid_recursion}
    In the grid network, consider a connected subgraph $S$. Suppose $|S| = j$ such that $j \leq \frac{3n}{4}$. Then, we have,
    \begin{align}
        v_S \leq \frac{\frac{2\lambda_e}{\lambda} + \lfloor \sqrt{j} \rfloor \max_{i \in N(S)}v_{S \cup \{i\}}}{\frac{j}{n} + \lfloor \sqrt{j} \rfloor} \label{grid_recursion_eq}
    \end{align}
\end{lemma}

\begin{Proof}
    We start with \eqref{rearranged_recursion},
    \begin{align}
       \lambda_e = \lambda_0(S)v_S + \sum_{i \in N(S)}\lambda_i(S)(v_S - v_{S \cup \{i\}}) \label{recur_eq}
    \end{align}
    Now, we define a function $E_S(i) = \sum_{k \in S} \mathbb{I}(\lambda_{ik} > 0)$, where $i \in N(S)$ and $\mathbb{I}(\cdot)$ is the indicator function. $E_S(i)$ is the number of incident edges on set $S$ that emanate at node $i$. Then, we can partition $N(S)$ as follows,
    \begin{align}
        A = \{i \in N(S): E_S(i) = 1\}\\
        B = \{i \in N(S): E_S(i) = 2\}\\
        C = \{i \in N(S): E_S(i) = 3\}\\
        D = \{i \in N(S): E_S(i) = 4\}
    \end{align}
    That is, $A$ is the set of neighbors of $S$ that update into the set $S$ with a single link (single edge), $B$ is the set of neighbors of $S$ that update into the set $S$ with two links (two edges), and so on. Using these sets, we bound \eqref{recur_eq} as,
    \begin{align}
        \lambda_e =& \lambda_0(S)v_S + \sum_{i \in A}\frac{\lambda}{4}(v_S - v_{S \cup \{i\}}) + \sum_{i \in B}\frac{\lambda}{2}(v_S - v_{S \cup \{i\}})\notag\\&+ \sum_{i \in C}\frac{3\lambda}{4}(v_S - v_{S \cup \{i\}}) + \sum_{i \in D}\lambda(v_S - v_{S \cup \{i\}})\\
        \geq &\lambda_0(S)v_S + |A|\frac{\lambda}{4}\min_{i \in A}(v_S - v_{S \cup \{i\}}) \notag\\
        &+ |B|\frac{\lambda}{2}\min_{i \in B}(v_S - v_{S \cup \{i\}}) \notag\\
        &+ |C|\frac{3\lambda}{4}\min_{i \in C}(v_S - v_{S \cup \{i\}}) \notag\\
        &+ |D|\lambda\min_{i \in D}(v_S - v_{S \cup \{i\}})\\
        \geq &\lambda_0(S)v_S + |A|\frac{\lambda}{4}\min_{i \in N(S)}(v_S - v_{S \cup \{i\}}) \notag\\&+ |B|\frac{\lambda}{2}\min_{i \in N(S)}(v_S - v_{S \cup \{i\}}) \notag\\
        &+ |C|\frac{3\lambda}{4}\min_{i \in N(S)}(v_S - v_{S \cup \{i\}}) \notag\\&+ |D|\lambda\min_{i \in N(S)}(v_S - v_{S \cup \{i\}})\\
        =& \lambda_0(S)v_S \notag\\
        &+ \frac{\lambda}{4}(|A| \!+\! 2|B| \!+\! 3|C| \!+\! 4|D|)\min_{i \in N(S)}(v_S \!-\! v_{S \cup \{i\}}) \!\!\\
        =& \lambda_0(S)v_S \notag\\
        &+ \frac{\lambda}{4}(|A| \!+\! 2|B| \!+\! 3|C| \!+\! 4|D|)(v_S \!-\! \max_{i \in N(S)}v_{S \cup \{i\}}) \!\!\label{eq_edgelb}
    \end{align}
    Now, we note that $E(S) = |A| + 2|B| + 3|C| + 4|D|$, where $E(S)$ is the total number of incoming edges into the set $S$. Thus, \eqref{eq_edgelb} states that for the two-dimensional grid, we have
    \begin{align}
     \lambda_e \geq \lambda_0(S)v_S + \frac{\lambda}{4} E(S) (v_S - \max_{i \in N(S)}v_{S \cup \{i\}}) \label{eq_edgelb_1}
    \end{align}
    
    Next, we use a result from \cite{harary1976extremal} and \cite{nowzari19improved} to provide a lower bound for the number of incoming edges into the set $S$. According to \cite{harary1976extremal}, on an infinite two-dimensional grid, of all the connected subsets with a fixed size $j$, the subset that has the minimum number of incoming edges into the subset is the \emph{spiral}. The number of incoming edges into a spiral of $j$ nodes is given by $2\lceil 2\sqrt{j} \rceil$. This bound has also been used in \cite{nowzari19improved} in the analysis of gossip algorithms for max-consensus on grids. Hence, we have the following lower bound for $E(S)$: $E(S) \geq 2\lceil 2\sqrt{j} \rceil \geq 4\lfloor \sqrt{j} \rfloor$. 

    \begin{figure}[t]
        \centering
        \includegraphics[width=\linewidth]{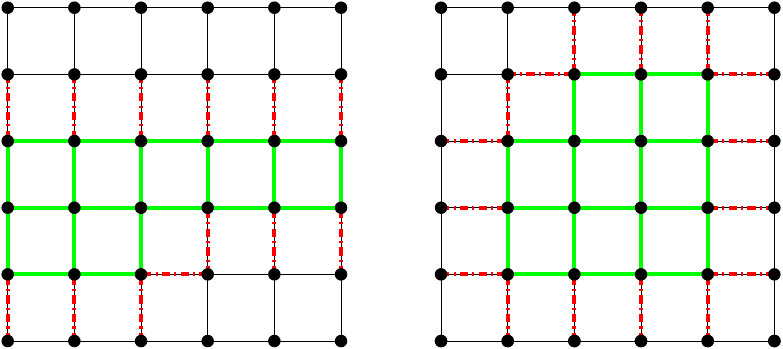}
        \caption{Comparison of the number of incoming edges of two sets with $15$ nodes on a $6\times 6$ grid network. Due to boundary constraints, the set on the left has less incoming edges than the spiral set on the right.}
        \label{fig_special_sets}
    \end{figure}
    
    However, our grid network is a finite network. Hence, the effects of the boundaries affect this bound once $j$ becomes large enough. To see this, consider the sets on a $6\times 6$ grid given in Fig.~\ref{fig_special_sets}. Both of the sets have $15$ nodes. The left set makes use of the boundaries in order to cut out the incoming edges on the extreme right and extreme left of the set. The right set is a spiral, and does not make use of the finite boundaries of the grid network. We note that the left set has $13$ incoming edges, whereas the spiral has $16$ incoming edges. Had the grid been infinite, the left set would have had $17$ incoming edges, one more than the spiral, satisfying the result in \cite{harary1976extremal}. We also observe that the set on the left has a lower bound on the number of incoming edges given by $2\sqrt{n}$. Hence, we conclude that up to a certain size of the set, the spiral will have lower number of incoming edges, and for sets larger than that will have the upper bound set by the set on the left in Fig.~\ref{fig_special_sets}. The exact inflection point is given by comparing the two bounds,
    \begin{align}
        2\sqrt{n} \geq 2\lceil 2\sqrt{j} \rceil
    \end{align}
    This is satisfied for $j < \frac{n}{4}$. Hence, we conclude that $E(S)$ is bounded below by $4\lfloor \sqrt{j} \rfloor$ up to $j = \frac{n}{4}-1$ and by $2\sqrt{n}$ for $j$ between $\frac{n}{4}$ and $\frac{3n}{4}$. Hence, considering both of these bounds, we use the common lower bound $E(S) \geq 2\lfloor \sqrt{j} \rfloor$ for all $j$ in our range of consideration. 
    Substituting this lower bound in \eqref{eq_edgelb_1}, we obtain
     \begin{align}
     \lambda_e \geq \lambda_0(S)v_S + \frac{\lambda}{4} 2\lfloor \sqrt{j} \rfloor (v_S - \max_{i \in N(S)}v_{S \cup \{i\}}) \label{eq_edgelb_2}
    \end{align}
    Inserting $\lambda_0(S)=\lambda j/n$ and rearranging \eqref{eq_edgelb_2}, we have
    \begin{align}
        v_S &\leq \frac{\lambda_e + \lambda \frac{\lfloor\sqrt{j}\rfloor}{2} \max_{i \in N(S)} v_{S \cup \{i\}}}{\frac{\lambda j}{n} + \lambda\frac{\lfloor\sqrt{j}\rfloor}{2}}\\
        &= \frac{\frac{2\lambda_e}{\lambda} + \lfloor \sqrt{j} \rfloor\max_{i \in N(S)} v_{S \cup \{i\}}}{\frac{2j}{n} + \lfloor \sqrt{j} \rfloor} \\
        &\leq \frac{\frac{2\lambda_e}{\lambda} + \lfloor \sqrt{j} \rfloor\max_{i \in N(S)} v_{S \cup \{i\}}}{\frac{j}{n} + \lfloor \sqrt{j} \rfloor}
    \end{align}
    which completes the proof.
\end{Proof}

This bound does not work for $j > \frac{3n}{4}$, since when $j$ increases beyond this, we have a new type of set with the least number of incoming edges on set $S$. This set $S$ is equivalent to the whole grid minus a spiral-shaped portion which has nodes that are not in $S$. The spiral has the minimum number of incoming edges, and hence the same can be said about the complement set of nodes. This results in set $S$ having the least number of incoming edges. The bound for the case $j > \frac{3n}{4}$ can be concluded from,
\begin{align} \label{eq_higher_number_of_nodes}
    2\lceil 2\sqrt{n-j}\rceil < 2\sqrt{n}
\end{align}
where $|S| = j$.

In order to find the version age of a single user, we use \eqref{grid_recursion_eq} recursively, and obtain an expression which contains floor functions. The floor functions are hard to analyze with regards to order scaling. Hence, we prove the following technical lemma, in order to get rid of the floor functions, and be able to carry out the analysis in Theorem~\ref{thm}.

\begin{lemma} \label{floor_inequality}
For any $n$, we have,
\begin{align}
    \sum_{i=1}^n \prod_{j=1}^i \frac{\lfloor \sqrt{j} \rfloor}{\lfloor \sqrt{j+1} \rfloor + \frac{j}{n}} \leq \sqrt{3}\sum_{i=1}^n \prod_{j=1}^i \frac{\sqrt{j}}{\sqrt{j+1} + \frac{j}{n}} \label{eq_lemma2}
\end{align}
\end{lemma}

\begin{Proof}
    First, we show that for $k \geq 2$, we have,
    \begin{align}
        \prod_{j=k^2-1}^{(k+1)^2-2} \frac{\lfloor \sqrt{j} \rfloor}{\lfloor \sqrt{j+1} \rfloor + \frac{j}{n}} \leq \prod_{j=k^2-1}^{(k+1)^2-2} \frac{\sqrt{j}}{\sqrt{j+1} + \frac{j}{n}} \label{eq_prod}
    \end{align}
    Firstly, we note that for $k^2-1< j \leq (k+1)^2-2$, $\lfloor \sqrt{j} \rfloor = \lfloor \sqrt{j+1} \rfloor = k$. For $j = k^2-1$, we have $\lfloor \sqrt{j} \rfloor = k-1$ and $\lfloor \sqrt{j+1} \rfloor = k$. Next, to investigate the validity of this fact, we upper bound the ratio of the left to right hand side expressions by $1$ using the following series of arguments,
    \begin{align}
        \!\!\prod_{j=k^2-1}^{(k+1)^2-2} &\frac{\frac{\lfloor \sqrt{j} \rfloor}{\lfloor \sqrt{j+1} \rfloor + \frac{j}{n}}}{\frac{\sqrt{j}}{\sqrt{j+1} + \frac{j}{n}}} = \left(\prod_{j=k^2}^{(k+1)^2-2} \frac{\frac{k}{k + \frac{j}{n}}}{\frac{1}{\sqrt{1+\frac{1}{j}} + \frac{\sqrt{j}}{n}}} \right)\frac{\frac{k-1}{k + \frac{k^2-1}{n}}}{\frac{\sqrt{k^2-1}}{k + \frac{k^2-1}{n}}} \! \label{eq_frac}\\
        &= \left(\prod_{j=0}^{2k-1} \frac{(1+\frac{1}{k^2+j})^{\frac{1}{2}} + \frac{\sqrt{k^2+j}}{n}}{1 + \frac{k}{n} + \frac{j}{kn}}\right) \sqrt{\frac{k-1}{k+1}} \label{eq_diff_before}\\
        &\leq \left(\frac{(1+\frac{1}{k^2})^{\frac{1}{2}} + \frac{k}{n}}{1 + \frac{k}{n}}\right)^{2k}\sqrt{\frac{k-1}{k+1}} \label{eq_diff_after}\\
        &= \left(1 + \frac{(1+\frac{1}{k^2})^{\frac{1}{2}} - 1}{1 + \frac{k}{n}}\right)^{2k}\sqrt{\frac{k-1}{k+1}}\\
        &\leq \left(1 + \left(\left(1+\frac{1}{k^2}\right)^{\frac{1}{2}} - 1\right)\right)^{2k}\sqrt{\frac{k-1}{k+1}}\\
        &= \left(1+\frac{1}{k^2}\right)^k\sqrt{\frac{k-1}{k+1}}
    \end{align}
    where we go from \eqref{eq_diff_before} to \eqref{eq_diff_after} by considering the function $g$ defined below for the range $k^2 \leq x < (k+1)^2-1$, $x \in \mathbb{R}$, where it is differentiable,
    \begin{align}
        g(x) = \frac{(1+\frac{1}{x})^{\frac{1}{2}} + \frac{\sqrt{x}}{n}}{1 + \frac{x}{kn}}
    \end{align}
    The derivative of $g$ is given by,
    \begin{align}
        g'(x) =& \frac{1}{2n^2kx\sqrt{x(x+1)}}( (k - 2x)x(x+1)^{\frac{1}{2}} \nonumber\\
        &+x\sqrt{x+1}(x-2n\sqrt{x+1}) - nx - k n^2)
    \end{align}
    which is always negative. Hence, the first term in the product in \eqref{eq_diff_before} dominates the rest of the term and then the inequality follows. Now, we want to show that the left hand side in \eqref{eq_frac} never exceeds $1$ for any $k \geq 2$. For this, we analyze the following continuous function,
    \begin{align}
        f(x) = 1 - \left(1+\frac{1}{x^2}\right)^x\sqrt{\frac{x-1}{x+1}}
    \end{align}
    We want to show that $f$ is a  strictly non-negative function for $x \geq 2$. To see this, we first write the derivative $f'$,
    \begin{align}
        f'(x) = \frac{(1+\frac{1}{x^2})^x(x^2-(x^4-1)\log{(\frac{1}{x^2}+1)}-3)}{\sqrt{\frac{x-1}{x+1}}(x+1)^2(x^2+1)}
    \end{align}
    The derivative $f'$ is always negative for $x \geq 2$, and its limit is $0$. Hence, we have that $f$ is a monotonically decreasing function with limit $0$ as $x$ tends to $\infty$. Thus, $f$ is always positive for finite values of $x$. Hence, we conclude that, 
    \begin{align}
        \prod_{j=k^2-1}^{(k+1)^2-2} \frac{\frac{\lfloor \sqrt{j} \rfloor}{\lfloor \sqrt{j+1} \rfloor + \frac{j}{n}}}{\frac{\sqrt{j}}{\sqrt{j+1} + \frac{j}{n}}} \leq 1
    \end{align}
    We also note that each term in the product in \eqref{eq_frac} is greater than $1$, whereas the only term outside the brackets is less than $1$. Hence, this one term dominates in the combined product, resulting in the product being less than $1$. Thus, we conclude that we can replace the upper limit in the product with any general number $i > k^2-1$. Finally, we substitute $k=2$ in \eqref{eq_prod}, along with the upper limit being a general $i$, we write,
    \begin{align}
        \prod_{j=3}^i \frac{\lfloor \sqrt{j} \rfloor}{\lfloor \sqrt{j+1} \rfloor + \frac{j}{n}} \leq \prod_{j=3}^i \frac{\sqrt{j}}{\sqrt{j+1} + \frac{j}{n}} \label{eq_3toi}
    \end{align}
    Note that the product of the first two terms on the left hand side of \eqref{eq_lemma2} are approximately $1$, and the same are $\frac{1}{\sqrt{3}}$ for the right hand side. Using this fact and \eqref{eq_3toi}, we finally write,
    \begin{align}
        \prod_{j=1}^i \frac{\lfloor \sqrt{j} \rfloor}{\lfloor \sqrt{j+1} \rfloor + \frac{j}{n}} \leq \sqrt{3}\prod_{j=1}^i \frac{\sqrt{j}}{\sqrt{j+1} + \frac{j}{n}}
    \end{align}
    Summing over all $i$ gives the desired result.
\end{Proof}

Next, we state and prove the main result of this paper.

\begin{theorem}\label{thm}
    The version age of a single user in the grid network scales as $O(n^{\frac{1}{3}})$.
\end{theorem}

\begin{Proof}
    We start with upper bounding the version age of a single user $v_1$ by a recursive application of \lemmaref{grid_recursion}. At each step, we can use this recursion because we do not care about the sets that are generated using the maximization at each step, since this path of growing the sets will lead to the final set being the full grid network, and we know that the version age of the full grid network is given by $v_{\mathcal{N}}=\frac{\lambda_e}{\lambda}$, irrespective of the sequence of sets that were generated at each step. We have two lower bounds of $E(S)$ for specific ranges of $j$ between $1$ and $n$. As we have discussed in Lemma~\ref{grid_recursion}, between $1$ and $\frac{3n}{4}$, the bound is $2\lfloor \sqrt{j}\rfloor$. The bound is $4\lfloor \sqrt{n-j}\rfloor$ between $\frac{3n}{4}$ and $n$, as discussed previously in \eqref{eq_higher_number_of_nodes}. 
    
    Now, we write the recursion, and separately bound the terms of the two ranges of $j$. First, we write a bound for the sum of terms $X$ corresponding to $j <\frac{3n}{4}$,
    \begin{align}
        X &\leq \frac{2\lambda_e}{\lambda}\left(\frac{1}{1+\frac{1}{n}}\right)\left(1 + \sum_{i=1}^{\frac{3n}{4}-1} \prod_{j=1}^i \frac{\lfloor \sqrt{j} \rfloor}{\lfloor \sqrt{j+1} \rfloor + \frac{j+1}{n}}\right) \!\!\label{eq_floor_before}\\ 
        &\leq \frac{2\lambda_e}{\lambda}\left(\frac{1}{1+\frac{1}{n}}\right)\left(1 + \alpha\sum_{i=1}^{\frac{3n}{4}-1} \prod_{j=1}^i \frac{\sqrt{j}}{\sqrt{j+1} + \frac{j}{n}}\right)\label{eq_floor_after}\\ 
        &\leq \frac{2\lambda_e}{\lambda}\left(1 + \alpha\sum_{i=1}^{\frac{3n}{4}-1} \prod_{j=1}^i \frac{\sqrt{j}}{\sqrt{j+1} + \frac{j}{n}}\right) \label{eq_approx}
    \end{align}
    where $\alpha = \sqrt{3}$. Here, we go from \eqref{eq_floor_before} to \eqref{eq_floor_after} by using \lemmaref{floor_inequality}. Next, to find the order of the right hand side of \eqref{eq_approx}, we proceed as follows,
    \begin{align}
        X &\leq \frac{2\lambda_e}{\lambda}\left(1 + \alpha\sum_{i=1}^{\frac{3n}{4}-1} \prod_{j=1}^i \frac{1}{(1+\frac{1}{j})^{\frac{1}{2}} + \frac{\sqrt{j}}{n}}\right) \label{eq_binom_approx_before}\\ 
        &\leq \frac{2\lambda_e}{\lambda}\left(1 + \alpha\sum_{i=1}^{\frac{3n}{4}-1} \prod_{j=1}^i \frac{1}{1 + \frac{1}{2}\frac{1}{j} - \frac{1}{8}\frac{1}{j^2} + \frac{\sqrt{j}}{n}}\right) \label{eq_binom_approx_after}
    \end{align}
    where \eqref{eq_binom_approx_after} uses the approximate binomial expansion to bound \eqref{eq_binom_approx_before}. Now, we define,
    \begin{align}
        a_i &= \prod_{j=1}^i \frac{1}{1 + \frac{1}{2}\frac{1}{j} - \frac{1}{8}\frac{1}{j^2} + \frac{\sqrt{j}}{n}} \label{def_ak}
    \end{align}
    By taking $\log$ of both sides, we obtain,
    \begin{align}
        -\log{a_i} &\approx \sum_{j=1}^i \left(\frac{1}{2}\frac{1}{j} - \frac{1}{8}\frac{1}{j^2} + \frac{\sqrt{j}}{n}\right)\\
        &\approx \frac{1}{2}(\log{i} + \gamma)- \delta + \frac{2i^{\frac{3}{2}}}{3n} \label{final_approx}
    \end{align}
    where $\gamma \approx 0.577$ is the Euler-Mascheroni constant, $\delta$ is a finite positive constant upper bounded by $\pi^2/48$ as $\sum_j \frac{1}{j^2} \rightarrow \frac{\pi^2}{6}$, and the third term is bounded by the first term of the Euler-MacLaurin series expansion. Substituting \eqref{final_approx} in \eqref{eq_approx},
    \begin{align}
        X &\leq \frac{2\lambda_e}{\lambda}\left(1 + \alpha'\sum_{i=1}^{\frac{3n}{4}-1} e^{-\frac{1}{2}\log{i} - \frac{2i^{\frac{3}{2}}}{3n}}\right)\\
        &= \frac{2\lambda_e}{\lambda}\left(1 + \alpha'\sum_{i=1}^{\frac{3n}{4}-1} \frac{1}{i^{\frac{1}{2}}} e^{-\frac{2i^{\frac{3}{2}}}{3n}}\right) \label{eq_reimann_sum}
    \end{align}
    We define a function $h$, 
    \begin{align}
        h(x) = \frac{1}{x^{\frac{1}{2}}}e^{-\frac{2}{3}x^{\frac{3}{2}}}
    \end{align}
    and then write the Riemann sum with step size $n^{\frac{2}{3}}$,
    \begin{align}
        \sum_{i=1}^{n} \frac{1}{n^{\frac{2}{3}}} h\left(\frac{i}{n^{\frac{2}{3}}}\right) = \int_{0}^{\infty} \frac{1}{t^{\frac{1}{2}}} e^{-\frac{2}{3}t^{\frac{3}{2}}}dt = \beta < \infty \label{beta}
    \end{align}
    Hence, we continue as follows,
    \begin{align}
        \sum_{i=1}^{n} \frac{1}{n^{\frac{2}{3}}} h\left(\frac{i}{n^{\frac{2}{3}}}\right) = \frac{1}{n^{\frac{1}{3}}} \sum_{i=1}^{n} \frac{1}{i^{\frac{1}{2}}} e^{-\frac{2i^{\frac{3}{2}}}{3n}} = \beta
    \end{align}
    Finally, continuing from \eqref{eq_reimann_sum}, we have,
    \begin{align}
        v_1 &\leq \frac{2\lambda_e}{\lambda}\left(1 + \alpha'\sum_{i=1}^{\frac{3n}{4}-1} \frac{1}{i^{\frac{1}{2}}} e^{-\frac{2i^{\frac{3}{2}}}{3n}}\right)\\
        &\leq \frac{2\lambda_e}{\lambda}\left(1 + \alpha'\sum_{i=1}^{n} \frac{1}{i^{\frac{1}{2}}} e^{-\frac{2i^{\frac{3}{2}}}{3n}}\right)\\
        &= \frac{2\lambda_e}{\lambda}(1 + \beta'n^{\frac{1}{3}}) \label{final_eq}
    \end{align}
    concluding that $X = O(n^{\frac{1}{3}})$.
    
    Next, we bound the terms that are corresponding to sets with sizes between $\frac{3n}{4}$ and $n$. We bound the sum $Y$ of these terms, using the fact that $E(S) \geq 4\lfloor\sqrt{n-j}\rfloor$, and hence (assuming $|S| = j$),
    \begin{align}
        v_S \leq \frac{\frac{\lambda_e}{\lambda} + \lfloor \sqrt{n-j} \rfloor \max_{i \in N(S)}v_{S \cup \{i\}}}{\frac{j}{n} + \lfloor \sqrt{n-j} \rfloor}
    \end{align}
    by a simple modification in Lemma~\ref{grid_recursion}. Continuing,
    \begin{align}
        Y \leq& \frac{\lambda_e}{\lambda}+\frac{\lambda_e}{\lambda}\left(1+\sum_{i=\frac{3n}{4}}^{n-2} \prod_{j=\frac{3n}{4}}^i \frac{\lfloor\sqrt{n-j}\rfloor}{\frac{j}{n} + \lfloor\sqrt{n-j-1}\rfloor}\right)\notag \\
        &\qquad \quad \times \prod_{j=1}^{\frac{3n}{4}-1}\frac{\lfloor\sqrt{j}\rfloor}{\frac{j}{n}+ \lfloor\sqrt{j+1}\rfloor} \times \frac{1}{\frac{3}{4} + \lfloor \frac{\sqrt{n}}{4}\rfloor}\\
        \leq& \frac{\lambda_e}{\lambda}+\frac{\lambda_e}{\lambda}\left(1+\sum_{i=\frac{3n}{4}}^{n-2} \prod_{k=n-i}^{\frac{n}{4}}\frac{\lfloor\sqrt{k}\rfloor}{\frac{n-k+1}{n}+ \lfloor\sqrt{k-1}\rfloor}\right)\notag\\
        &\qquad \quad \times \prod_{j=1}^{\frac{3n}{4}-1}\frac{\lfloor\sqrt{j}\rfloor}{\lfloor\sqrt{j+1}\rfloor} \times \frac{1}{\lfloor \frac{\sqrt{n}}{2}\rfloor}\\
        \leq& \frac{\lambda_e}{\lambda}+\frac{\lambda_e}{\lambda}{\frac{1}{\lfloor\frac{\sqrt{n}}{2}\rfloor}}\frac{1}{\lfloor\sqrt{\frac{3n}{4}}\rfloor}\left(1+\sum_{i=\frac{3n}{4}}^{n-2} \prod_{k=n-i}^{\frac{n}{4}}\frac{\lfloor\sqrt{k}\rfloor}{\lfloor\sqrt{k-1}\rfloor}\right)\\
        =&  \frac{\lambda_e}{\lambda}+\frac{\lambda_e}{\lambda}\left(1+\frac{1}{\lfloor\sqrt{\frac{3n}{4}}\rfloor}\sum_{i=\frac{3n}{4}}^{n-2} \frac{1}{\lfloor\sqrt{n-i-1}\rfloor}\right) \\
        =& \frac{\lambda_e}{\lambda}+\frac{\lambda_e}{\lambda}\left(1+\frac{1}{\lfloor\sqrt{\frac{3n}{4}}\rfloor}\sum_{l=1}^{\frac{n}{4}-1} \frac{1}{\lfloor\sqrt{l}\rfloor}\right)\\
        \leq& \frac{\lambda_e}{\lambda}+\frac{\lambda_e}{\lambda}\left(1+\frac{1}{\sqrt{\frac{3n}{4}}-1}\sum_{l=1}^{\frac{n}{4}-1} \frac{1}{\sqrt{l}-1}\right)\\
        \leq& \frac{\lambda_e}{\lambda}+\frac{\lambda_e}{\lambda}\left(1+\sum_{l=1}^{\frac{n}{4}-1} \frac{1}{l}\right)\\
        \leq& \frac{\lambda_e}{\lambda}(2+\nu \log{n})\label{later_terms}
    \end{align}
    where $\nu$ is a constant. Hence, using \eqref{final_eq} and \eqref{later_terms}, we write,
    \begin{align}
        v_1 &\leq X + Y\\
        &= \frac{2\lambda_e}{\lambda}(1 + \beta'n^{\frac{1}{3}})+\frac{\lambda_e}{\lambda}(2+\nu \log{n})
    \end{align}
    hence concluding that $v_1 = O(n^{\frac{1}{3}})$
\end{Proof}

\section{Discussion}

\begin{remark}
    In \eqref{beta}, $\beta = (\frac{2}{3})^{2/3}\Gamma(\frac{1}{3})$, where $\Gamma(\cdot)$ is the gamma function. Thus, $\beta' = \sqrt{3} \times e^{-\frac{\gamma}{2}}\times e^{\pi^2/48} \times (\frac{2}{3})^{2/3}\Gamma(\frac{1}{3}) \approx 3.2594$. Hence, as $n \rightarrow \infty$, the second term in \eqref{final_eq} dominates, and the version age of a single node is upper bounded as,
    \begin{align}
        v_1 \leq 6.5188\frac{\lambda_e}{\lambda}n^{\frac{1}{3}} \label{54}
    \end{align}
\end{remark}

\begin{remark} \label{remark2}
    As an interesting research direction, the results of this work can be generalized to a general $d$-dimensional grid from the two-dimensional case of this work. In a $d$-dimensional grid, each node has $2d$ neighbors. For this extension, we need generalizations of the results in \cite{harary1976extremal} and \cite{nowzari19improved} for $d$-dimensional grids, and prove Lemma~\ref{floor_inequality} after replacing $j^{\frac{1}{2}}$ by $j^\frac{d-1}{d}$ in this case. After these generalizations, calculations similar to those in Theorem~\ref{thm} will show that the version age of a single node in a $d$-dimensional grid scales as $O(n^{\frac{1}{d+1}})$. This result seems to be true by simulations. It is also satisfied by the three known cases: In the disconnected case $d=0$ and the version age is $O(n)$, in the ring/line network $d=1$ and the version age is $O(n^{\frac{1}{2}})$, and in the two-dimensional grid case (this paper) $d=2$ and the version age is $O(n^{\frac{1}{3}})$.
\end{remark}

\begin{remark}
    We note that for the disconnected network the version age scales as $O(n)$, for the ring/line network the version age scales as $O(n^{\frac{1}{2}})$, for the two-dimensional grid network the version age scales as $O(n^{\frac{1}{3}})$. If the conjecture in Remark~\ref{remark2} is correct, then the version age for a $d$-dimensional grid scales as $O(n^{\frac{1}{d+1}})$. Even though these results show the improvement in version age as the connectivity increases, these results still do not describe the entire spectrum of connectivity going from the fully-disconnected network to the fully-connected network where the version age scales as $O(\log n)$, since the nodes in the $d$-dimensional grid have $2d$ neighbors, which is still a constant even though it increases with $d$. In contrast, in the fully-connected network, each node has $n-1$ neighbors, which depends on the number of nodes $n$ in the network.
\end{remark}

\begin{remark}
There are many works considering gossip networks with the version age metric in different settings. These include adversarial models such as timestomping \cite{kaswan22timestomp} and jamming \cite{kaswan22jamming}, hierarchical structures such as version age with a community structure \cite{buyukates21CommunityStructure}, extended gossip algorithms such as file slicing and network coding \cite{kaswan22slicingcoding}, and opportunistic gossiping in multiple access networks \cite{mitra_allerton22,mitra2023timely}. All these papers use either the ring network or the fully-connected network as the system model. With the results of the present paper, extension of these results to grid networks is possible.  
\end{remark}

\begin{remark}
    \begin{figure}[t]
        \centering
        \includegraphics[width = 0.7\linewidth]{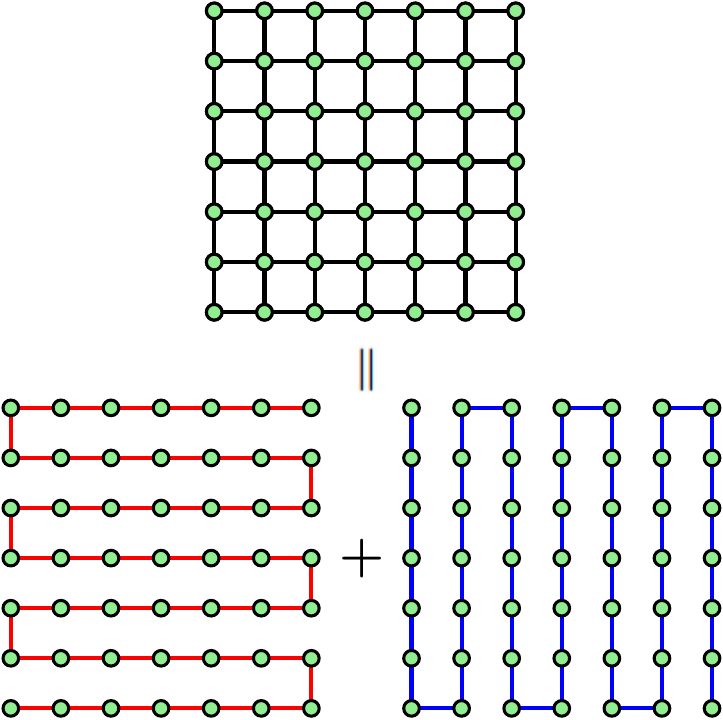}
        \caption{The grid can be thought of as two conjoined line networks, which reduces the version age from $O(n^{\frac{1}{2}})$ to $O(n^{\frac{1}{3}})$.}
        \label{fig_conjoined_grid}
    \end{figure}
    It is well known that conjoining two networks disseminating the same information speeds up the information diffusion \cite{yagan2013conjoining}. For example, suppose there is a group of friends living in the same city, who meet each other regularly. They want to be up-to-date about the happenings in each other's lives. Then, we see that they will receive updates about their friends faster if they were connected to each other on a social media platform and also met each other, rather than doing just one of the two. This is an interesting perspective with which we can also look at the grid network as a conjoined network of two rings/lines as shown in \figref{fig_conjoined_grid}, which improves the version age of the network from $O(n^{\frac{1}{2}})$ to $O(n^{\frac{1}{3}})$.    
\end{remark}

\begin{remark}
    Reference \cite{buyukates21CommunityStructure} considers gossip networks with a community structure. It considers a system where a source node receives an update and disseminates it to cluster heads, which in turn disseminate this information to cluster nodes under them. It is shown in \cite{buyukates21CommunityStructure} that having a community structure improves the version age for several community structures. These community (i.e., cluster) structures include the disconnected network, the ring network and the fully-connected network. Following \secref{sec3_1}, we can upper bound the recursion found in \cite[Thm.~1]{buyukates21CommunityStructure} to obtain,
    \begin{align}
        v_S \leq &\frac{\lambda_e + \lambda_c(S)v_c}{\lambda_c(S) + |N(S|\min_{i \in N(S)}\lambda_i(S)}\notag\\ 
        &+ \frac{|N(S)|\min_{i \in N(S)}\lambda_i(S)\max_{i \in N(S)}v_{S \cup \{i\}}}{\lambda_c(S) + |N(S|\min_{i \in N(S)}\lambda_i(S)}
    \end{align}
    where there are $m$ communities each with $k$ nodes such that $n = mk$ and $v_c = m\lambda_s/\lambda_e$. Then, assuming $\lambda_c = \lambda$ and that clusters are grid networks as we have defined in this paper, we can upper bound $v_1$ in one cluster of the community as,
    \begin{align}
        v_1 \leq v_c(1 - a_{k}) + 2\beta'\frac{\lambda_e}{\lambda}k^{\frac{1}{3}}
    \end{align}
    where $a_k$ is as defined in \eqref{def_ak}. Now, as $n \rightarrow \infty$, $a_k \rightarrow 0$. Hence, we have,
    \begin{align}
        v_1 \leq m\frac{\lambda_e}{\lambda_s} + 2\beta'\frac{\lambda_e}{\lambda}k^{\frac{1}{3}}
    \end{align}
    Choosing $m = n^\frac{1}{4}$ and $k = n^\frac{3}{4}$, by adding community structure, we can improve the version age from $O(n^{\frac{1}{3}})$ to $O(n^{\frac{1}{4}})$. Assuming that Remark~\ref{remark2} is correct, we can carry out a similar exercise for a $d$-dimensional grid and conclude that we can improve the version age from $O(n^{\frac{1}{d+1}})$ to $O(n^{\frac{1}{d+2}})$ by introducing community structure to the network.
\end{remark}

\begin{figure}[t]
    \centering
    \includegraphics[width = \linewidth]{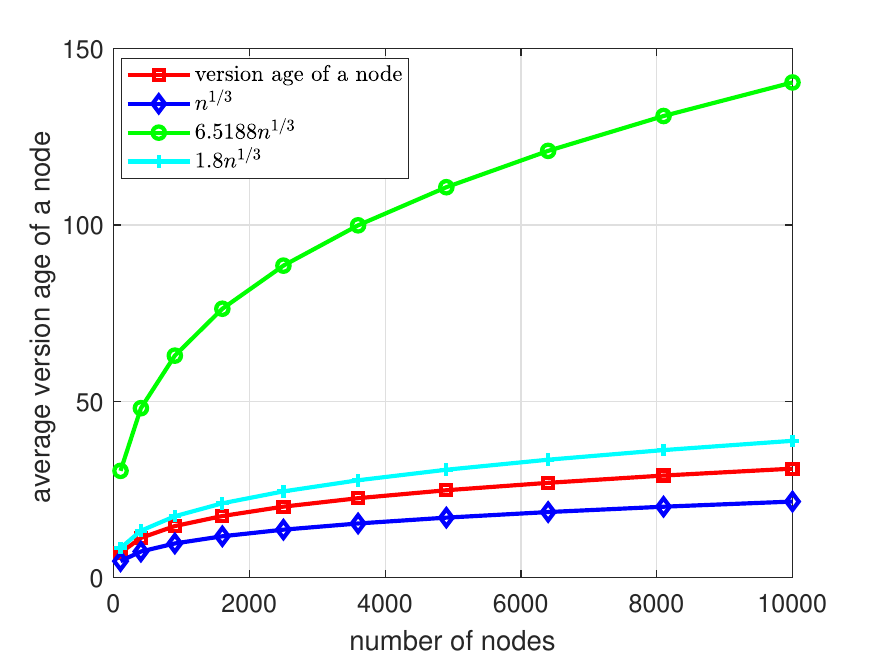}
    \caption{Version age of a single node versus the number of nodes in a two-dimensional grid gossip network.}
    \label{fig_grid_order}
\end{figure}

\section{Numerical Results} \label{sec_numerical_results}

We conduct numerical simulations to find the version age of a single node in the two-dimensional grid network by setting $\lambda_e = 1$, $\lambda = 1$ and varying $n$ from $100$ nodes, corresponding to a $10 \times 10$ grid, to $10,000$ nodes, corresponding to a $100 \times 100$ grid, such that the number of nodes on one edge of the grid increases by increments of $10$. We plot the results in \figref{fig_grid_order}. The simulated version age is plotted along with the upper bound $6.5188n^{\frac{1}{3}}$ we obtained in \eqref{54}. We see by simulations that our calculated upper bound bounds the simulated version age. We also plot $1.8n^{\frac{1}{3}}$ and $n^{\frac{1}{3}}$ to provide closer upper and lower bounds for the simulated version age. By trying various multiples for $n^{\frac{1}{3}}$ to match with the simulated version age, we find that $1.45n^{\frac{1}{3}}$ is a close approximation to the simulated version age we have obtained.

\section{Conclusion}

We considered a two-dimensional square grid in which each node receives information from a source node and through gossip from its neighbors. We used the version age of information metric to quantify the freshness of information of each node. We found a general upper bound for the average version age of a subset of the network. Specializing this bound to the two-dimensional grid network, we showed that the average version age of a single node in a grid scales as $O(n^{\frac{1}{3}})$ as a function of the network size $n$.

\bibliographystyle{unsrt}
\bibliography{refs}

\end{document}